\documentclass[preprint,5p,twocolumn,12pt]{elsarticle}

\usepackage{amssymb,amsmath}

\newcommand{\reff}[1]{(\ref{#1})}

\newcommand{\x}{\ol{x}}

\newcommand{\av}[1]{\left|#1\right|}

\newcommand{\brkts}[1]{\left(#1\right)}
\newcommand{\ebrkts}[1]{\left[#1\right]}

\newcommand{\brcs}[1]{\left\{#1\right\}}

\newcommand{\bsplitl}[2]{
\begin{equation}
\begin{split}
#1
\end{split}
\label{#2}
\end{equation}}

\journal{Physics Letters A}

\begin{document}

\begin{frontmatter}



\title{New and general framework for adsorption processes on dynamic interfaces}


\author[ChemE,Math]{Markus Schmuck}
\author[ChemE]{Serafim Kalliadasis\corref{cor1}}
\ead{s.kalliadasis@imperial.ac.uk}
\cortext[cor1]{Corresponding author.}
\address[ChemE]{Department of Chemical Engineering,
Imperial College London, London SW7 2AZ, United Kingdom}
\address[Math]{Department of Mathematics,
Imperial College London, London SW7 2AZ, United Kingdom}

\begin{abstract}
We introduce a new and general continuum thermodynamic framework for the
mathematical analysis and computation of adsorption on dynamic interfaces. To
the best of our knowledge, there is no formulation available that accounts
for the coupled dynamics of interfaces and densities of adsorbants. Our
framework leads to analytic adsorption isotherms which also take the
interfacial geometry fully into account. We demonstrate the utility and
physical consistency of our framework with a new computational multi-level
discretization strategy. In the computations, we recover the experimentally
observed feature that the adsorption of particles minimizes the interfacial
tension.
\end{abstract}


\end{frontmatter}


\section{Introduction}
\label{sec:1}
Adsorption processes on interfaces are of great importance in a wide spectrum
of scientific and industrial applications such as the stabilization of foams
\cite{Aveyard2003}, the formulation of nanoporous materials, functional
membranes and capsules, drug delivery \cite{Koch2011,Yang2011} and oil
recovery \cite{Sun2011}, to name but a few. Also, many aerated food products
such as bread cake, meringue, ice-cream and mousse are based on stabilized
emulsions. Not surprisingly, (particle) adsorption has been an active topic
of both experimental and theoretical research for several decades. Equally
important are mixtures of different types of particles which require a
dynamic description of the interfacial tension due to interactions between
the different species involved, especially if one considers surfactants. In
general, the interface between any two bulk neighbouring phases characterizes
and mediates any physical process occuring between these two phases.
Additional complexities include adsorption phenomena observed for charged
particles \cite{Danov2006}, capping ligands \cite{Gar2012}, or swelling latex
particles \cite{Maxwell1992} on fluid-fluid interfaces.

Here, we develop a new and generic free-energy formalism for dynamic but
diffuse interfaces in arbitrary spatial dimensions. A small number of
previous modelling attempts was restricted to a specific one-dimensional
setting with a fixed interface so far~\cite{Ariel1999,Diamant1996}. We
incorporate the experimentally important characterization of particles by the
contact angle into our mathematical description. Our derivation is based on
well-established and rigorous mathematical results which connect the diffuse
interface approach, i.e., the van der Waals/Cahn-Hilliard free-energy
formalism, to the interfacial tension. This result is known as the
Modica-Mortola theorem \cite{Modica1977} and relates the interfacial tension
to the homogeneous free energy, i.e., the classical and phenomenological
double-well potential $W(s):=(1-s^2)^2/4$.

We investigate in detail the case of ad- and de-sorption of uniform particles \cite{Garbin2012b}
and then give a generalization towards mixtures of particles which even allow to account for
surface active species, such as surfactants \cite{Nagarajan1991,Diamant}. We further extend the new free
energy formulation towards incompressible and immiscible fluid flows. Our
approach also provides a solid basis for extensions of studies on the
formation and evolution of electrical double layers in electrochemistry
\cite{Plieth2008,Newman2010B}, with applications in energy storage devices
such as supercapacitors, batteries, and micro-fluidic devices.

\section{Adsorption on dynamic interfaces}
\label{sec:2}
In one dimension and for a fixed fluid-fluid interface at $x=0$, the
following free energy
\bsplitl{
\gamma^{ex}(c)
	:= \int_0^\infty f_b^{ex}(c)\,dx
	+f_I(c)\,,
}{1DFiIf}
was studied by Andelman, Ariel, and Diamant
\cite{Ariel1999,Diamant1996} and by Mohrbach \cite{Mohrbach2005a}. The
variable $f_b^{ex}$ stands for the excess bulk free energy which is defined
by the ideal entropy of mixing, i.e. $f_b^{ex}(c):=\brkts{k_BT\ebrkts{c({\rm
log}\,c-1)-c_b({\rm log}\,c_b-1)}-\mu_b(c-c_b)}/a^3$, where $a>0$ is the
molecular dimension of the particles. To properly account for the higher
particle concentration at a fluid-fluid interface, the non-ideal entropy of
mixing with an additional adsorption parameter $\alpha$, and interaction
parameter $\beta$ is introduced by
\bsplitl{
f_I(c)
	& := \Bigl(k_BT\ebrkts{
			c{\rm ln}\,c
			+(1-c){\rm ln}(1-c)
		}
\\&\quad
	-\alpha c
	-\beta/2 c^2
	-\mu_1 c
	\Bigr)/a^2\,,
}{fI}
where $\mu_1=\mu_b:=k_BT{\rm ln}\,c_b$ is equal to the bulk chemical
potential. This simplified formulation does not account for the interface
formed between the two immiscible fluids \cite{Liu2003}. Here, we shall
develop a generic theoretical and computational framework that accounts for
both the particle densities and a dynamic immiscible interface which are
nonlinearly coupled. Our framework builds also the basis for studying
specific particles such as surfactants via the interaction energy defined via
$\beta$ in \reff{fI}. First, we include the characterization of particles by
the contact angle into the free energy \reff{fI}. Based on the well-known
free-energy change induced by adsorbing a single particle at the fluid-fluid
interface \cite{Aveyard2003,Gar2012,Pieranski1980}, we correspondingly write
for a change in particle density \bsplitl{ \delta F_{ads}
	:=-\gamma_{12}\pi \brkts{1-\av{\cos\,\theta}}^2\overline{\delta c} \,,
}{Fads} where $\theta$ is the contact angle formed at the triple phase
interface. Via a first-order Taylor approximation around equilibrium at the
interface $I$, i.e., $\gamma^{ex}(c_b+\overline{\delta c})\Bigr|_{I}\approx
\gamma^{ex}(c_b)\Bigr|_{I} +\frac{\delta\gamma^{ex}(c_b)}{\delta
c}\overline{\delta c}\Bigr|_{I}$, we can identify \bsplitl{
\delta F_{ads}
	= \brkts{ \gamma^{ex}(c_b+\overline{\delta c}) - \gamma^{ex}(c_b) }\Bigr|_{I}
	= \frac{\delta\gamma^{ex}}{\delta c}(c_b)\Bigr|_{I}\overline{\delta c}
	\,, }{padsorb} where $\overline{\delta c}$ is a small variation of the
interfacial particle density. This finally allows us to define $\alpha$ for
an interaction energy $\beta$ given as the minimum potential energy of an
associated inter-particle potential, e.g. of Lennard-Jones type.

\subsection{Asorption of uniform particles}
\label{sec:2_1}
We now generalize \reff{1DFiIf} towards a diffuse interface formulation by
introducing an additional order parameter $\phi$ motivated by the
well-accepted Cahn-Hilliard approximation, i.e.,
$e(\phi):=W(\phi)+\lambda^2\av{\nabla\phi}^2$, where $W(s):=(1-s^2)^2/4$ is
the dimensionless double-well potential. We note that $\lambda$ is
proportional to the interfacial width and hence proportional to the molecular
dimension $a$ and the particle concentration $c$.
Variational considerations imply that $W(\phi)$ is only non-zero in a diffuse
neighborhood around the interface. Therefore, we define the following
free-energy density \bsplitl{ f(\phi,c)
	:=e(\phi)+
	\sigma(\lambda)W(\phi)f_I(c)\,, }{fphi} where
$\sigma(\lambda):=3\sqrt{2}/\lambda$ (see also~\cite{Marc2012,Markus2012}).
From \cite{Elliott2010}, one can approximate the characteristic function for
the interface by $\chi_I(\phi)\approx\frac{3\sqrt{2}}{\lambda}W(\phi)$. The
definition \reff{fphi} is physically and mathematically motivated by the
Modica-Mortola theorem \cite{Modica1977} which relates the surface tension to
the double-well potential $W$. Moreover, the energy density $e$ in
\reff{fphi} is known to represent the perimeter of the phase enclosed by the
interface in the sharp interface limit $\lambda\to 0$. Finally, via the
approximate characteristic function of the phase $\phi=+1$, i.e.,
$\chi_1(\phi):=(\phi+1)/2$, we can introduce the new interfacial energy
(i.e., interfacial tension which depends on the particle density) for
arbitrary, dynamic, and higher dimensional interfaces by \bsplitl{
\overline{\gamma}^{ex}(\phi,c)
	:= \int_\Omega f(\phi,c)+\chi_1(\phi)f_b^{ex}(c)\,d{\bf x}
	\,.
}{Ge1}
The first variation of \reff{Ge1} with respect to $c$ leads to the following chemical potential at the
interface $I$,
\bsplitl{
\mu_c^{ex}\Bigl|_{I}
	:= \frac{\delta\overline{\gamma}^{ex}}{\delta c}\Bigl|_{I}
	& = \Bigl(
		k_BT{\rm log}\frac{c}{1-c}
		-\alpha
		-\beta c
\\&\quad
		-\mu_1
	\Bigr)
		\sigma(\lambda)W(\phi)
	\,.
}{mu1}
For the equilibrium bulk potential $\mu_1=\mu_b=k_BT{\rm log}\,c_b$, we obtain with \reff{mu1}
the new adsorption isotherm
\bsplitl{
c
	= \frac{c_b}{
		c_b
		+{\rm exp}\ebrkts{
			-(\alpha+\beta c)
				\sigma(\lambda)W(\phi)
			/(k_BT)
		}
	}\,,
}{GeAdIs}
which also accounts for the state of the interior phase. If we neglect
interactions, i.e., $\beta=0$, and set $\sigma(\lambda)=k_BT/W(\phi)$, then
\reff{GeAdIs} represents the classical Langmuir adsorption isotherm. Figure 1
depicts the adsorption isotherm \reff{GeAdIs} for an interface at $x=0$ and
the well-known equilibrium profile $\phi_{eq}(x) =
\tanh\brkts{x/(2\sqrt{2}\lambda)}$. The adsorption on a circular interface is
shown in Fig.~2. Under these conditions and after setting \reff{mu1} to zero,
we further obtain an expression for the equilibrium surface pressure
$\Pi(\phi_{eq},c_{eq})=-\overline{\gamma}^{ex}(\phi,c)\Bigr|_{eq}$.

\begin{figure}[t]
\includegraphics[width=4.2cm]{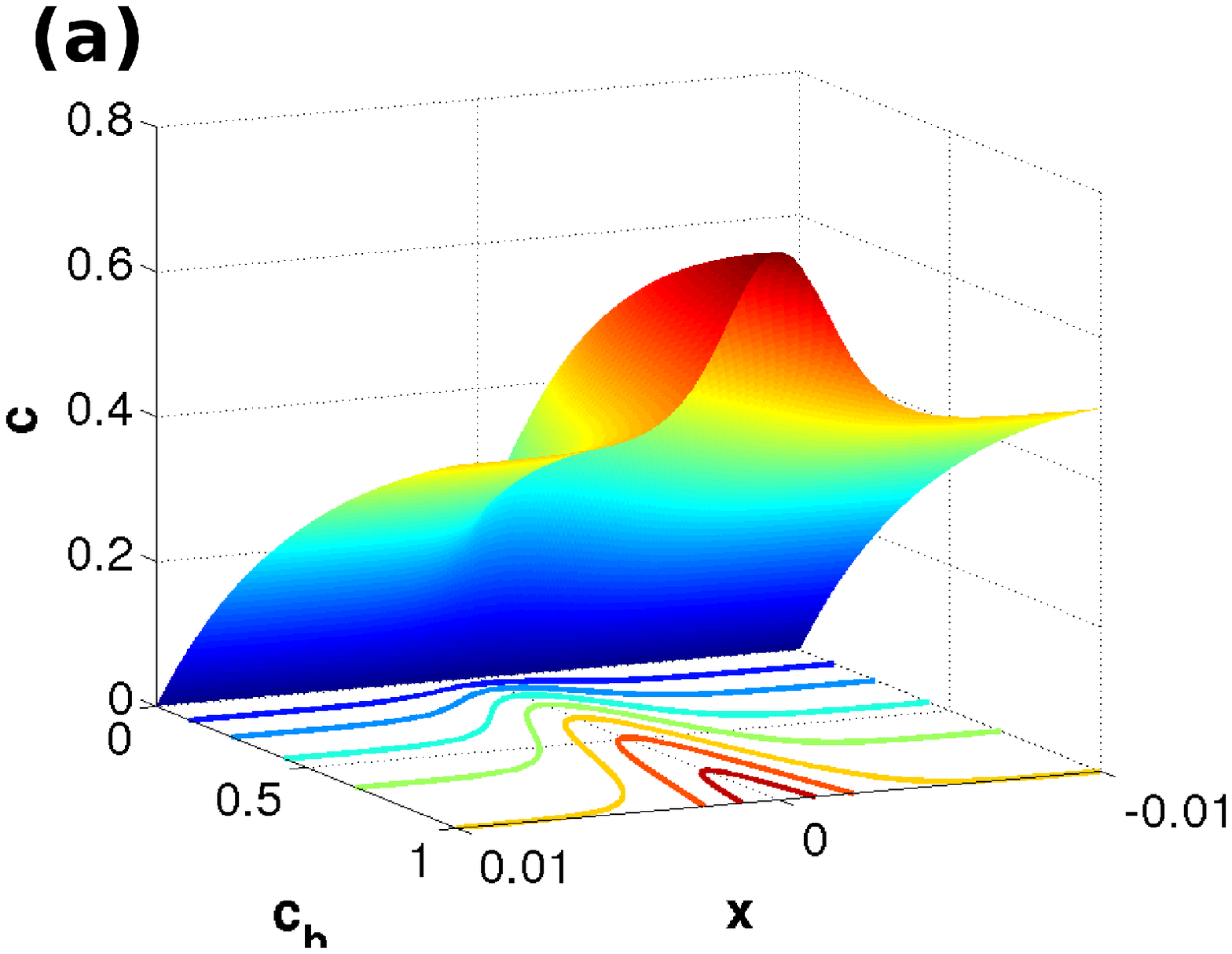}
\hfill
\includegraphics[width=4.2cm]{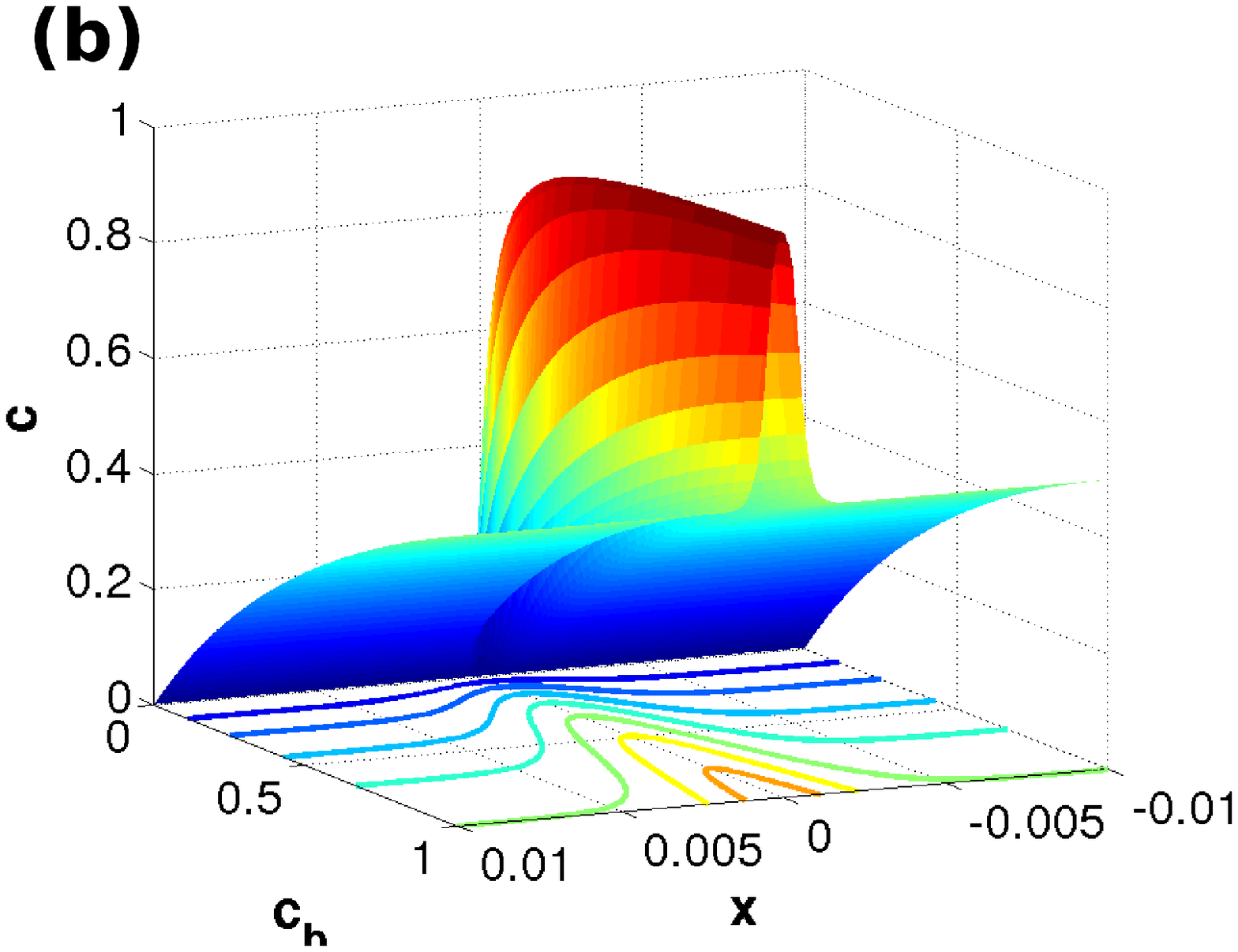}
\caption{Generalized adsorption isotherm in \reff{GeAdIs} for $\phi_{eq}(x) = \tanh\brkts{x/(2\sqrt{2}\lambda)}$ and
different interfacial widths $\lambda$. {\bf (a)} $\alpha=3.0$, $\lambda=0.0005$. {\bf (b)} $\alpha=10.0$, $\lambda=0.00015$.
}
\label{fig1}
\end{figure}


The expression in \reff{Ge1} can be interpreted as the free energy of the
interface $\phi$ with particle concentrations $c$. We then derive evolution
equations by the principle of steepest descent and its functional
generalization to gradient flows. That means, we obtain the gradient flows
\bsplitl{
\begin{cases}
\quad
\frac{\partial\phi}{\partial t}
	= {\rm div}\brkts{
		M_\phi\nabla\brkts{
			\frac{\delta\overline{\gamma}^{ex}}{\delta\phi}
		}
	}\,,
	&
\\\quad
\frac{\partial c}{\partial t}
	= {\rm div}\brkts{
		M_c\nabla\brkts{
			\frac{\delta\overline{\gamma}^{ex}}{\delta c}
		}
	}\,,
	&
\end{cases}
}{GrFl1} where the equations are made dimensionless by the bulk diffusion
time scales $\tau_\phi:=\frac{k_BT a^3}{M_\phi}$ and
$\tau_c:=\frac{k_BTa^3}{M_c}$. We note that the scalar mobilities $M_{\phi}$
and $M_c$ can be related to a scalar product with respect to which we measure
the variational derivatives $\frac{\delta \gamma^{ex}}{\delta\phi}$ and
$\frac{\delta \gamma^{ex}}{\delta\phi}$. It is noteworthy that a new
stochastic mode reduction strategy for gradient-flow systems was recently
proposed in~\cite{Schmuck2013} and it allows for a rigorous dimensionally
reduced description for such systems.


\subsection{Asorption of different types of particles}
\label{sec:2_2}
For simplicity, we state subsequent equations for identical molecular
dimension $a$ for both types of particles $c_1$ and $c_2$ while noting that a
generalization towards different molecular dimensions is straightforward. In
order to account for mixtures of two (i.e., $c_1$ and $c_2$, whereas an
according extension to an arbitrary but finite number of different types of
particles is straightforward), we extend the interfacial free energy
\reff{fI} as follows \bsplitl{ & f_I(c_1,c_2,\phi)
	=
	\frac{1}{a^2}\biggl(
		c_1{\rm ln}\,c_1
		+c_2{\rm ln}\,c_2
		+c_3{\rm ln}\,c_3
		-\epsilon_{12} c_1c_2
\\&\qquad
		-(\alpha_1+\mu_1)c_1
		-(\alpha_2+\mu_2)c_2
		-\frac{\beta_1}{2}c_1^2
		-\frac{\beta_2}{2}c_2^2
	\biggr)
\,,
}{fI2}
where $c_3:= 1-c_1-c_2$ is the surface coverage of solvent represented by the order parameter $\phi$ (e.g. water/oil) and $\epsilon_{12}$ is the interaction
energy between $c_1$ and $c_2$.
Based on \reff{fphi} and \reff{fI2}, we then define the generalized surface tension $\overline{\gamma}^{ex}$
according to \reff{Ge1}. Herewith, we obtain the two chemical potentials
\bsplitl{
\mu_1^{ex}\bigr|_I
	& :=
		\frac{\delta\overline{\gamma}^{ex}}{\delta c_1}
	= \Bigl(
		k_BT{\rm log}\frac{c_1}{c_3}
		-\alpha_1-\mu_1
\\&
		-\beta_1 c_1
		-\epsilon_{12}c_1
	\Bigr)
		\sigma(\lambda)W(\phi)		
	\,,
\\
\mu_2^{ex}\bigr|_I
	& :=
		\frac{\delta\overline{\gamma}^{ex}}{\delta c_2}
	= \Bigl(
		k_BT{\rm log}\frac{c_2}{c_3}
		-\alpha_2-\mu_2
\\&
		-\beta_2 c_2
		-\epsilon_{12}c_2
	\Bigr)
		\sigma(\lambda)W(\phi)		
	\,.
}{mu1mu2}
At equilibrium, the excess chemical potentials \reff{mu1mu2} are set to zero by the
assumption that the chemical potential is globally equal to the bulk values. This leads then to the
adsorption isotherms,
\bsplitl{
c_1
	= \frac{
		c_{1b}(1-c_2)
	}{
		c_{1b}+{\rm e}^{
			-\brkts{
				\alpha_1+\beta_1c_1+\epsilon_{12}c_2
			}
				\sigma(\lambda)W(\phi)/(k_BT)
		}
	}\,,
\\
c_2
	= \frac{
		c_{2b}(1-c_1)
	}{
		c_{2b}+{\rm e}^{
			-\brkts{
				\alpha_2+\beta_2c_2+\epsilon_{12}c_1
			}
			\sigma(\lambda)W(\phi)/(k_BT)
		}
	}
\,,
}{GeAdIs2}
where $c_{1b}$ and $c_{2b}$ are the equilibrium bulk densities.
If we neglect the particle interactions via $\beta_i$, $i=1,2$, and 
$\epsilon_{12}$, then we again end up with the Langmuir adsorption 
isotherm.


\begin{figure}
\includegraphics[width=4.0cm]{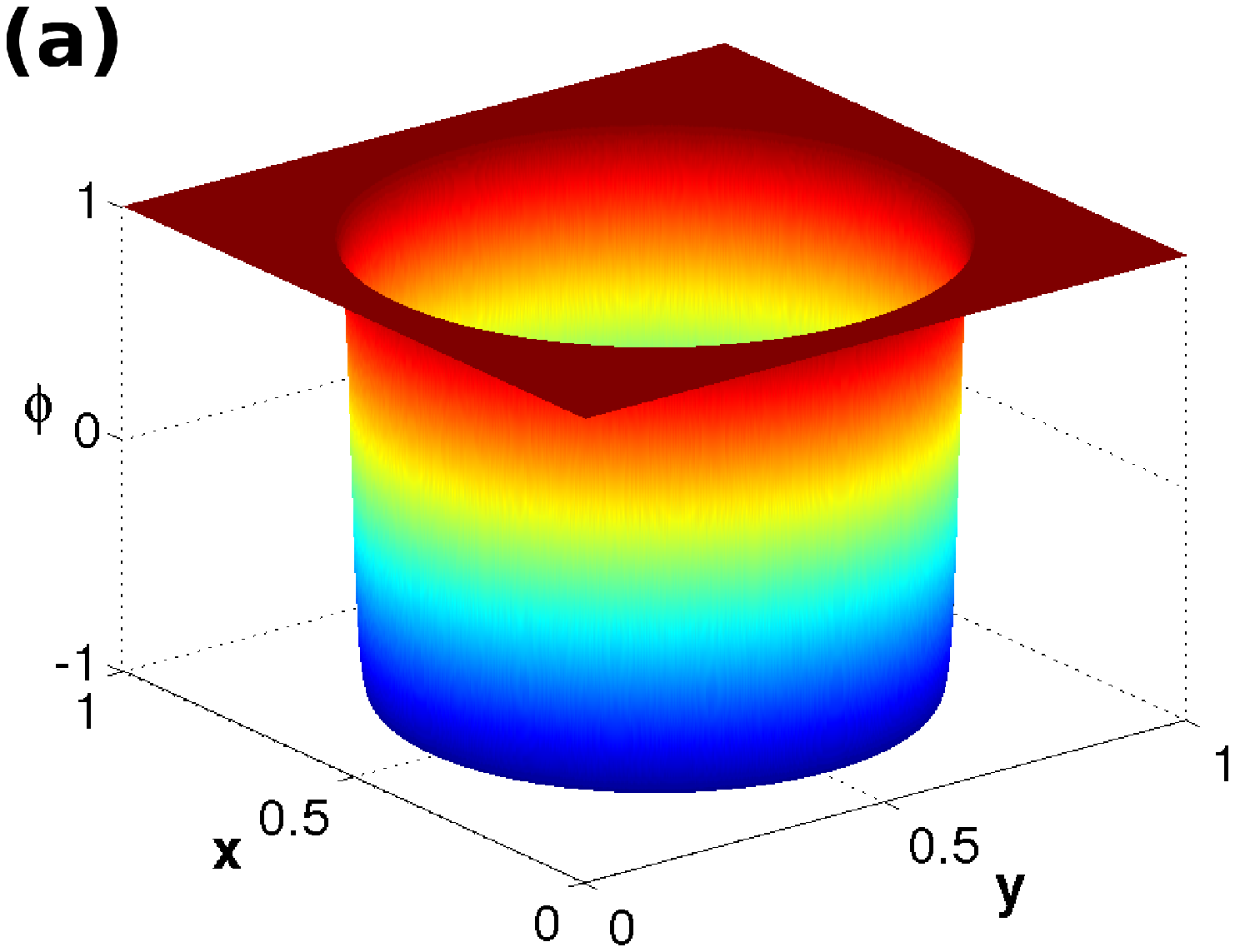}
\hfill
\includegraphics[width=4.2cm]{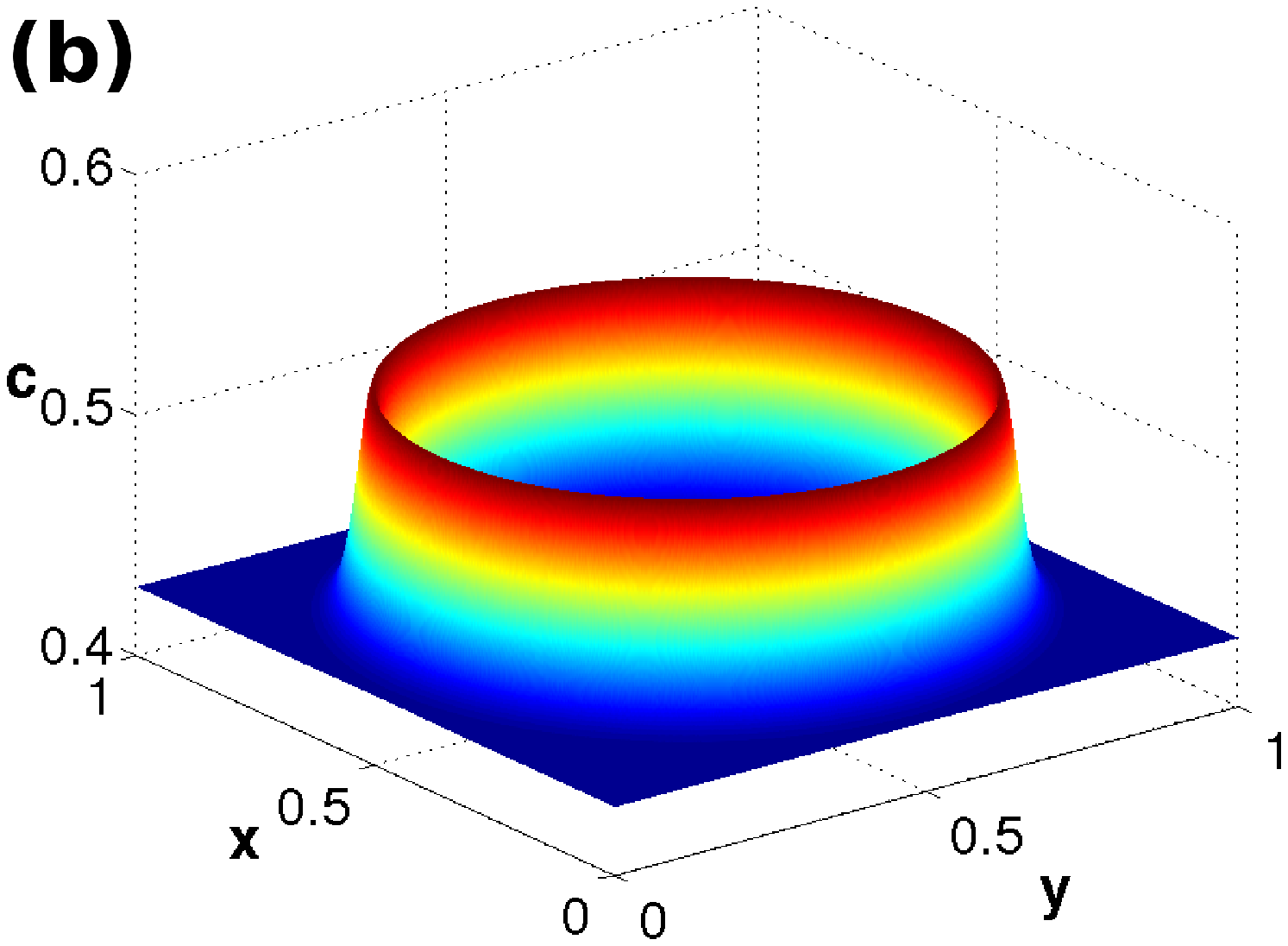}
\\\center
\includegraphics[width=4.3cm]{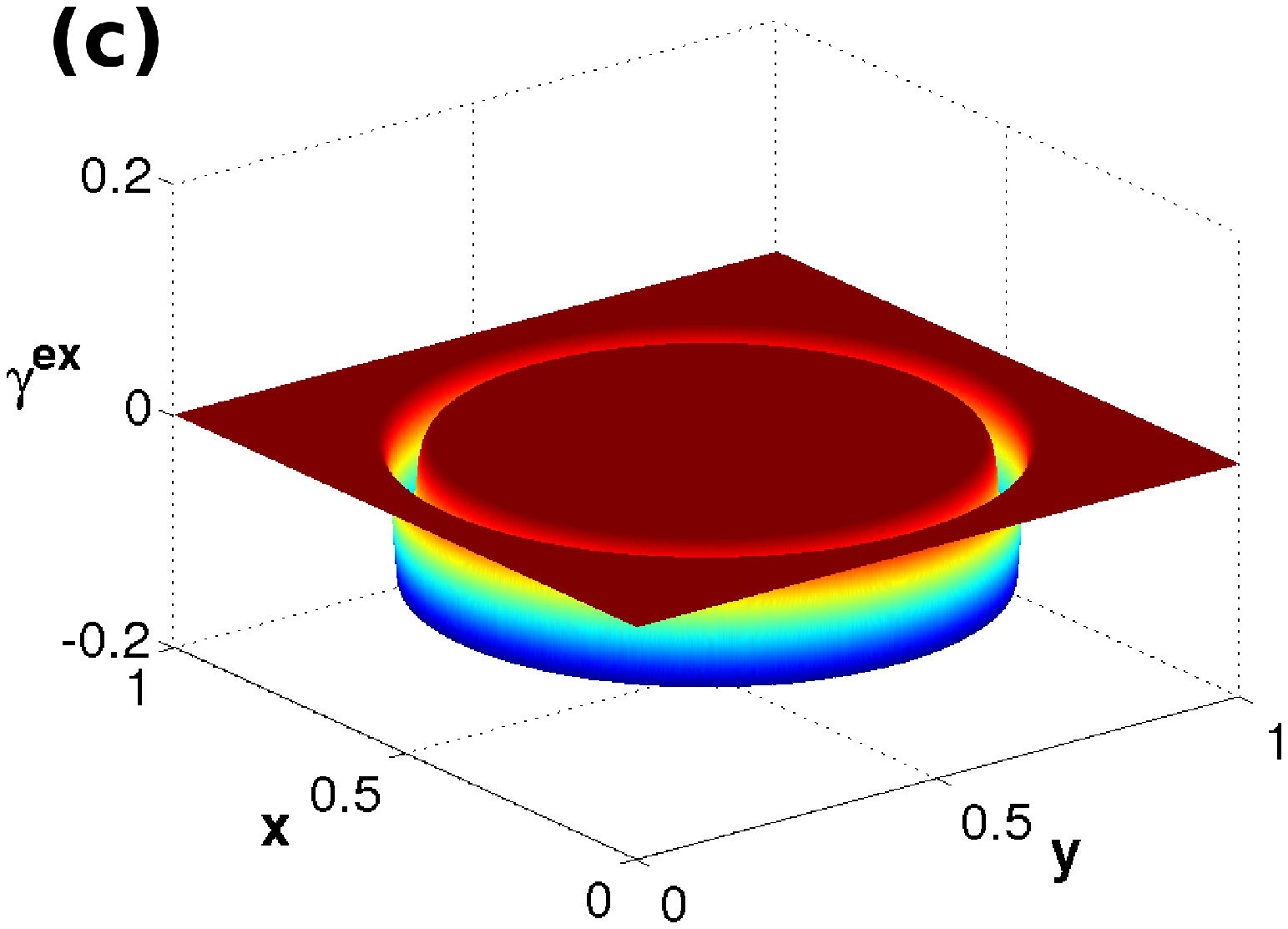}
\caption{{\bf (a)} Equilibrium phase field variable $\phi_{eq}(x) = \tanh\brkts{d_r(x)/(2\sqrt{2}\lambda)}$ defined via the signed
distance function $d_r(x)=\sqrt{(x-x_0)^2+(y-y_0)^2}-r$. {\bf (b)} Generalized adsortion isotherm in equation (\ref{GeAdIs}) with
$c_b=0.75$, $\lambda=0.0005$, and $r=0.4$. {\bf (c)} Free energy $\gamma^{ex}(c_{eq},\phi_{eq})$ for equilibrium
particle concentration which is computed under an equilibrium phase field $\phi_{eq}$.
}
\label{fig2}
\end{figure}

We also recover the same dynamic equations \reff{GrFl1} 
after replacing \reff{GrFl1}$_2$ by two corresponding equations, i.e., for the chemical potentials \reff{mu1mu2} instead of
$\frac{\delta\overline{\gamma}^{ex}}{\delta c}$ introduced in \reff{mu1mu2}.

\subsection{Generalization to incompressible fluids}
\label{sec:2_3}
We adapt the variational principles in \cite{Liu2003,deGennes1993} in order
to generalize (a) and (b) to include fluid flow. We extend
$\overline{\gamma}^{ex}$ to the action functional \bsplitl{ & F({\bf x}({\bf
X},t)
		= \int_0^T\int_\Omega\biggl(
		\frac{1}{2}\av{{\bf x}_t({\bf X},t)}^2
\\&
		-\frac{\lambda^2}{2}\av{\hat{\rm D}^{-1}\nabla_{\bf x}\phi({\bf x}({\bf X},t),t)}^2
		-\lambda^2W(\phi({\bf x}({\bf X},t),t))
\\&
		-\sigma(\lambda)W(\phi({\bf x}({\bf X},t),t))f_I(c({\bf x}({\bf X},t),t))
\\&
		-\chi_1(\phi({\bf x}({\bf X},t),t))f_b^{ex}(c({\bf x}({\bf X},t),t))
	\biggr){\rm det}\,\hat{\rm D}\,d{\bf X}\,dt
\,,
}{AP}
for the interfacial width $\lambda\propto (\textrm{surface tension})\times(\textrm{capillary width})$.
We restrict ourselves to incompressible fluids and define the reference (Eulerian) coordinate
${\bf x}({\bf X},t)$ via the material (Lagrangian) coordinate ${\bf X}$, i.e.,
\bsplitl{
{\bf x}_t({\bf X},t)
	= {\bf u}({\bf x}({\bf X},t),t)\,,
\qquad
{\bf x}({\bf X},0)
	= {\bf X}\,,
}{FlMa}
and
$
\hat{\rm D}({\bf x}({\bf X},t),t)
	=\frac{\partial {\bf x}({\bf X},t)}{\partial {\bf X}}
$
is th deformation tensor (strain) of the flow map [\ref{FlMa}]. The kinetic
energy density $\frac{1}{2}\av{{\bf x}_t({\bf X},t)}^2$ leads to the Eulerian
part of the incompressible Navier-Stokes equation. The terms arising due to
the presence of particles, are $W'(\phi)f_I(c)\nabla\phi$ and
$W(\phi)f_I'(c)\nabla c$ such that we end up with the following momentum
equation \bsplitl{
\begin{cases}
\quad
\frac{\partial{\bf u}}{\partial t}
	+{\bf u}\cdot\nabla{\bf u}
	-\Delta{\bf u}
	+\nabla p
	&
\\\qquad
	=
	-\lambda^2{\rm div}\brkts{
		\nabla\phi\otimes\nabla\phi
		-W(\phi)\hat{\rm I}
	}&
\\\qquad
	- \sigma(\lambda)W'(\phi)f_I(c)\nabla\phi
	- \sigma(\lambda)W(\phi)f_I'(c)\nabla c
	\,,&
\\\quad
{\rm div}\brkts{{\bf u}}
	= 0\,,&
\end{cases}
}{MoEq} where $\hat{\rm I}$ is the identity tensor. We note that the viscous
term $-\Delta{\bf u}$ in equation [\ref{MoEq}] enters by the maximum
dissipation principle (PMD) \cite{Svoboda2006a,Hackl2008,Hyon2010}. The
variational framework \reff{AP} and \reff{FlMa} also extends the gradient
flow \reff{GrFl1} for the single and mixed particles to fluid flow by the
coupled partial differential equations (PDEs) \bsplitl{
\begin{cases}
\quad
\frac{\partial\phi}{\partial t}
	+ {\bf u}\cdot\nabla\phi
	= {\rm div}\brkts{
		M_\phi\nabla\brkts{
			\frac{\delta F}{\delta\phi}
		}
	}\,,
	&
\\\quad
\frac{\partial c}{\partial t}
		+ {\bf u}\cdot\nabla c
	= {\rm div}\brkts{
		M_c\nabla\brkts{
			\frac{\delta F}{\delta c}
		}
	}\,.
	&
\end{cases}
}{PaEq}

\section{Computations: Heterogeneous multi-level method}
\label{sec:3}
Here, we advocate a new heterogeneous multi-level (HML) approach which
combines the physical ideas from statistical thermodynamics and probabilistic
transport theory in a new stochastic multiscale scheme. The framework
consists of two basic steps:

\vspace{0.25cm}

\emph{(1) Macro-level discretization/interpolation:} We introduce a Galerkin discretization in order to resolve differential operators such
as gradients in the domain ${\cal D}$ by interpolation over nodal values with associated coordinates $\pmb{\eta}\in {\cal N}_h$ where ${\cal N}_h$
denotes the set of all nodes of a quasi-uniform triangulation ${\cal T}_h$ of ${\cal D}$ into triangles
or tetrahedrons of maximal diameter $h>0$. Herewith, we can introduce the
affine finite element space
$
V_h
	=\brcs{
		\Phi\in C(\overline{\Omega})\,:\,\Phi\,|_{K}\in P_1(K), K\in {\cal T}_h
	}
$
and an associated nodal interpolation ${\cal I}_h\,:\, C(\overline{\cal D})\,\to\,V_h$ such that
${\cal I}_hu({\bf x})=\sum_{\pmb{\eta}\in {\cal N}_h}u(\pmb{\eta})\varphi_{\pmb{\eta}}$ where $\brcs{\varphi_{\pmb{\eta}}\,:\,\pmb{\eta}\in{\cal N}_h}\subset V_h$
denotes the nodal basis for $V_h$ and $u\in C(\overline{\cal D}).$

\emph{(2) Micro-level computation:} A systematic way to compute the macroscopic nodal values $\brcs{u(\pmb{\eta})}_{\pmb{\eta}\in {\cal N}_h}$ is to rewrite the PDE for the particles as a Brownian motion (with reflection in the case of Neumann boundary
conditions) starting at each node $\pmb{\eta}\in{\cal N}_h$ and stopped as soon as the Brownian walker hits the Dirichlet boundary for the
first time. If there is no Dirichlet boundary, one needs to define a large enough time horizon. The expectation of the Brownian paths starting at the nodal coordinates $\pmb{\eta}\in{\cal N}_h$ then defines the nodal values $\brcs{u(\pmb{\eta})}_{\pmb{\eta}\in {\cal N}_h}$ where
$u$ for example stands for the particle concentration $c$.

The development of a rigorous basis for stochastic approximations which
converge with order 1 with respect to the time discretization parameter for
Laplace and Poisson equations has recently found much interest
\cite{Bossy2004,Lejay2013}. However, the convergence rate of our
computational strategy depends additionally on the interpolation error on the
macro level since the schemes studied in \cite{Bossy2004,Lejay2013} analyze
only the computation of a single nodal value. For simplicity, we subsequently
compute random walks on a lattice $\mathbb{ Z}^d$ where we move randomly
single particles and accept the new state by a Metropolis acceptance
criterium. Since we need mass conservation for the particles, we employ
Kawasaki-type dynamics.

\vspace{0.25cm}

The advantage of the multi-level computational strategy developed here is
that it allows the computation of particle trajectories in an efficient and
systematic fashion. In fact, an additional strength of our strategy is that
it allows for a direct parallelization in the sense that one can compute the
random walks for each macroscopic node $\pmb{\eta}\in{\cal N}_h$ at the same
time on different cores of a processor. Alternatively, often finite element methods
are applied requiring adaptive meshes for
feasible computations \cite{Banas2008,Stogner2006}.

\begin{figure}[t]
\includegraphics[width=.2\textwidth]{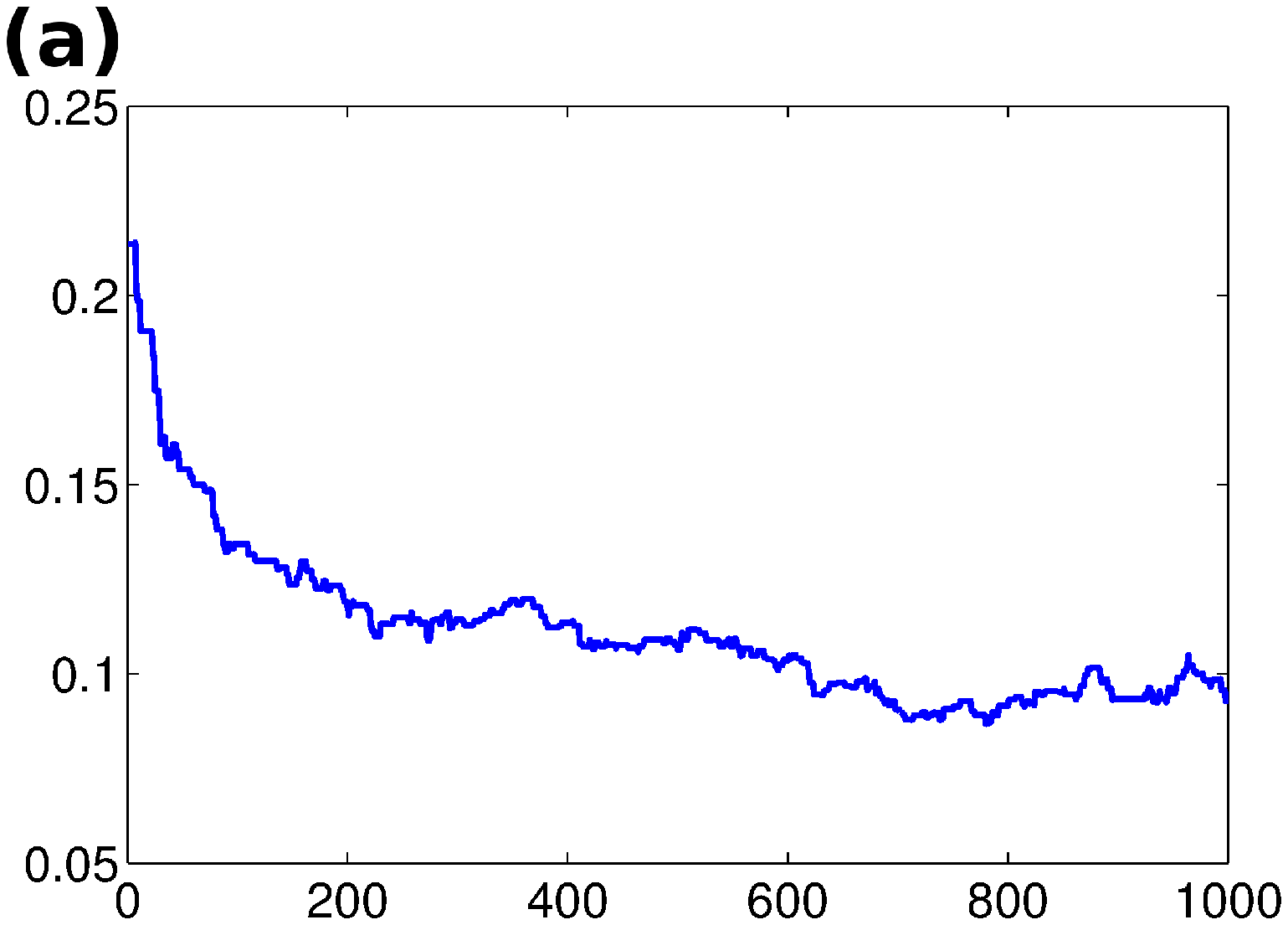}
\hfill
\includegraphics[width=.17\textwidth]{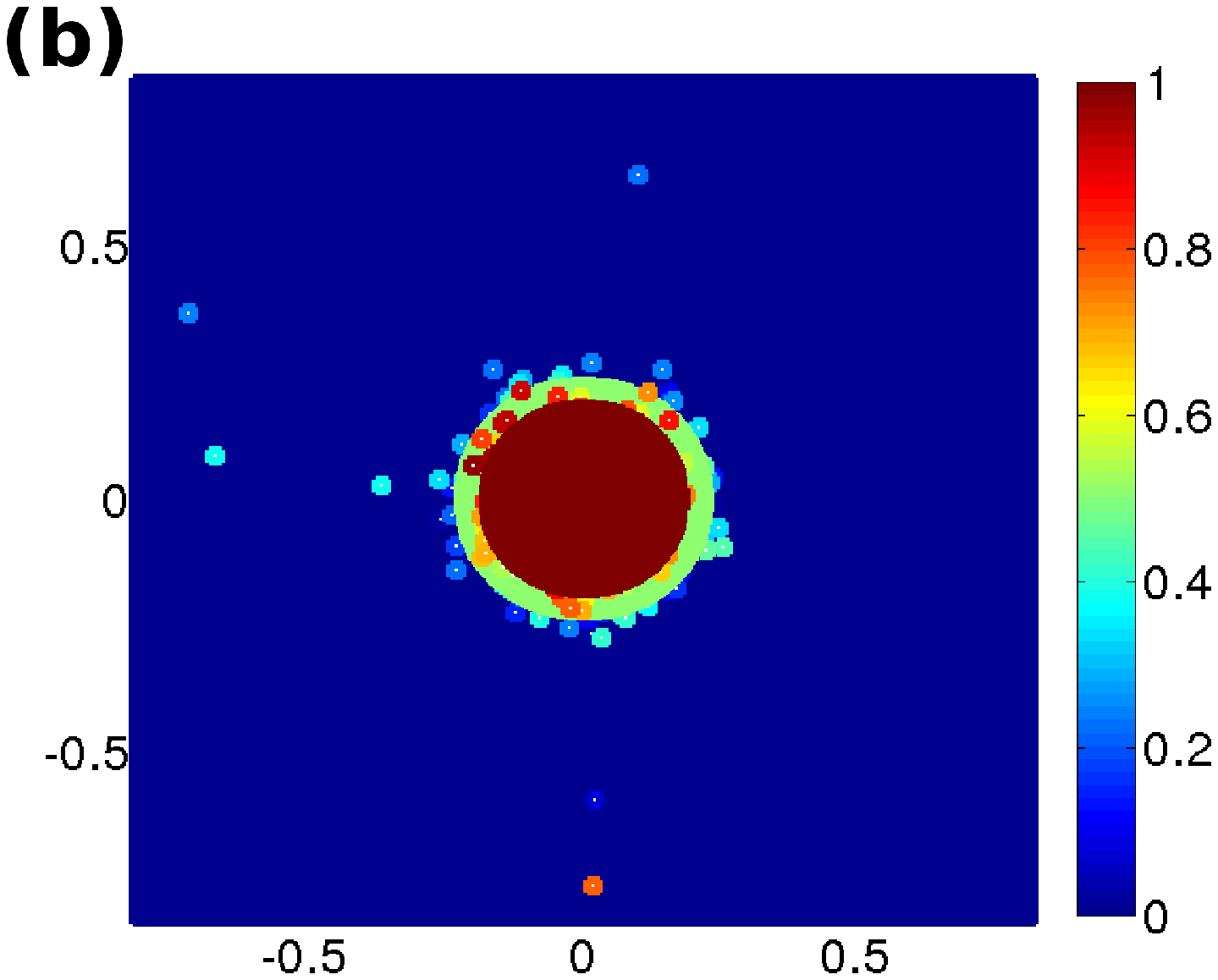}
\\
\includegraphics[width=4cm]{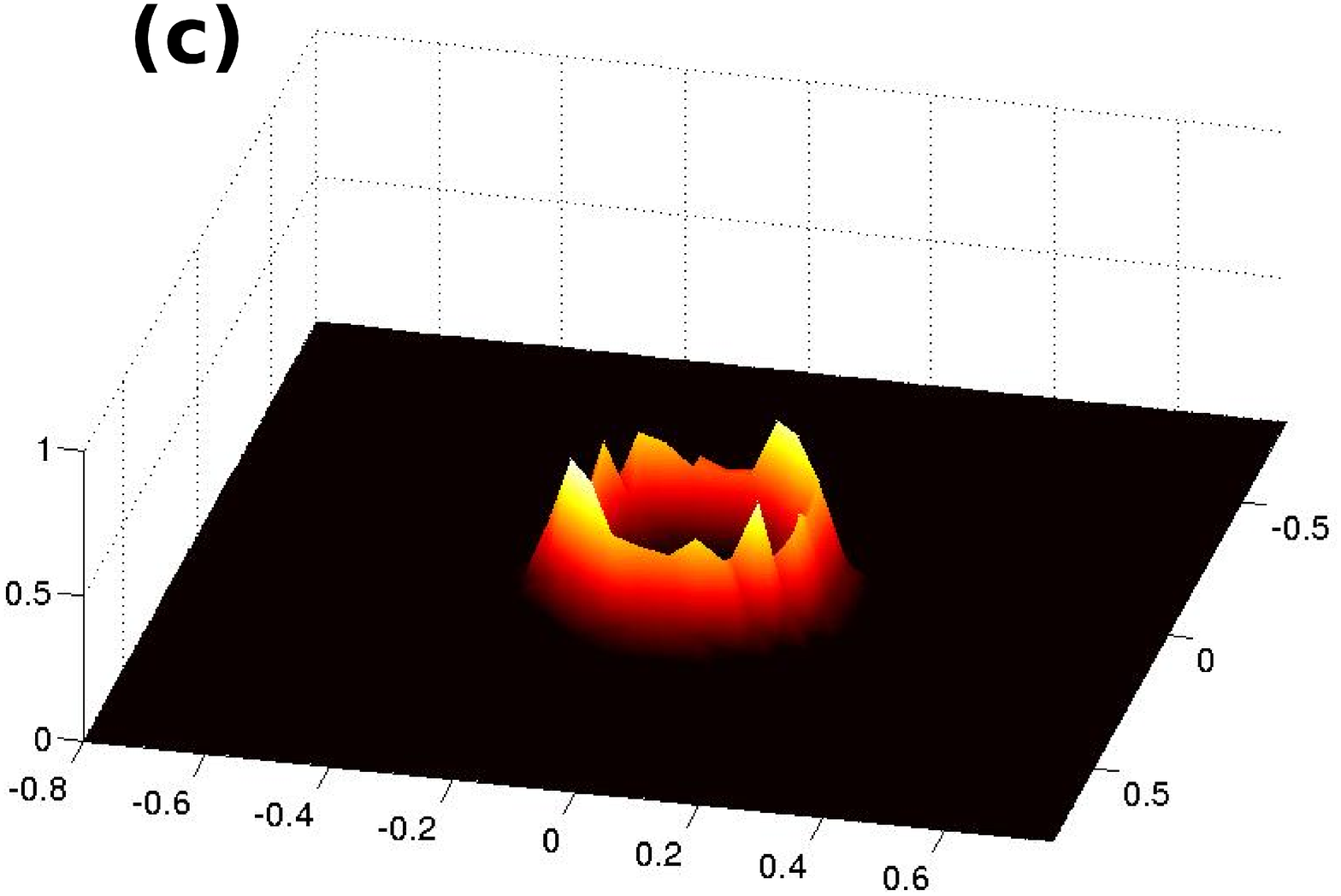} 
\hfill
\includegraphics[width=3cm]{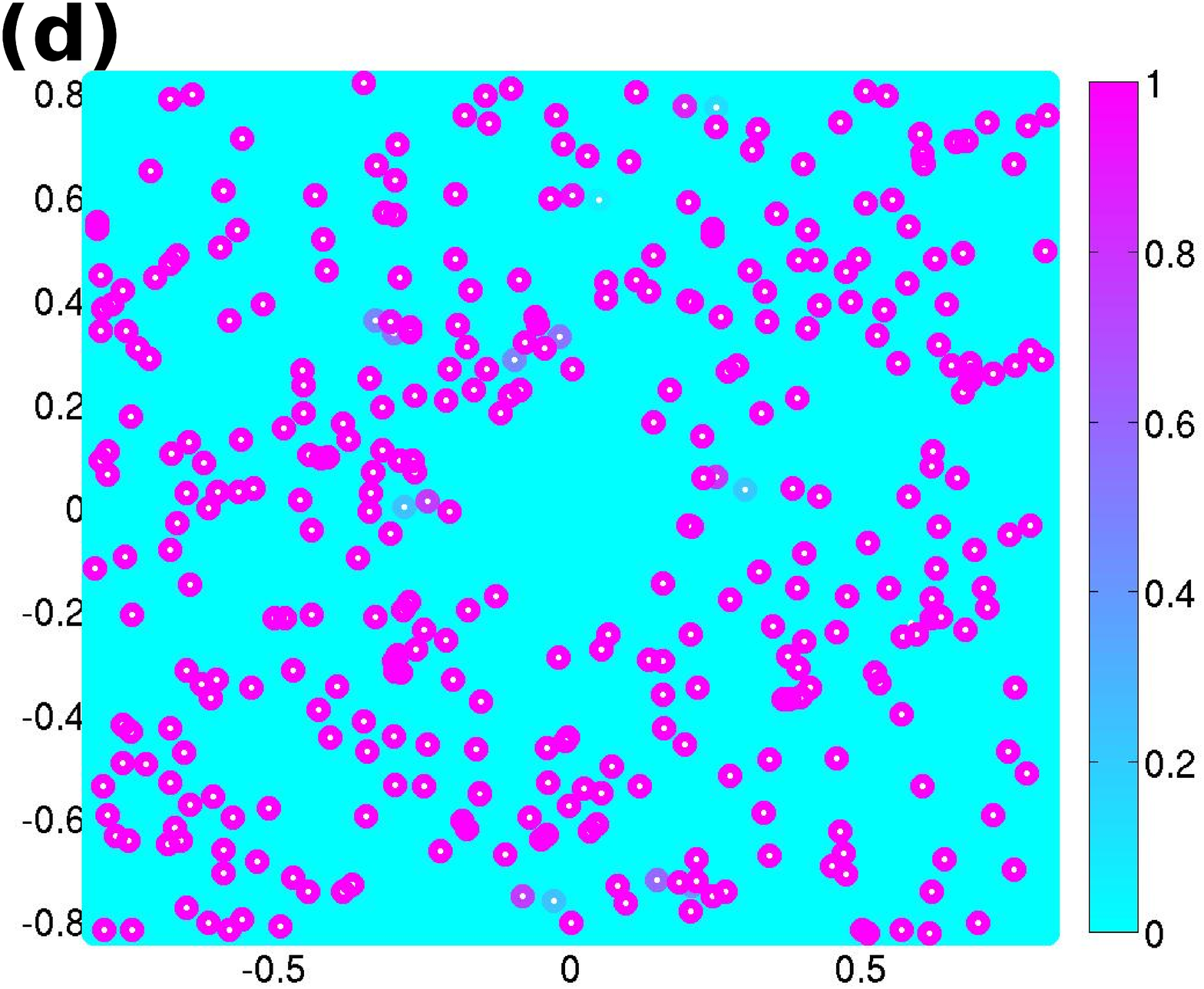} 
\caption{{\bf (a)} Decay of interfacial tension as a result of minimizing the free energy \reff{Ge1}. {\bf (b)} Microscopic dynamics of the phase field $\phi$.
{\bf (c)} Interfacial energy density. {\bf (d)} 275 particles diffusing in phase $\phi=0$.
}
\label{figPhQ2}
\end{figure}

Subsequently, we scrutinize our modelling and computational framework with
two model computations in which fluid flow is neglected for simplicity. In
the first example, we perform computations with 275 particles diffusing in
one phase of a two-phase immiscible fluid. Our domain is the square
$[-1,1]^2$ which is discretized by a macroscopic mesh of size $h_1=0.04$ for
the first level and the second microscopic level is computed on a grid of
size $h_2=h_1/14$. In Fig.~3 we depict characteristic quantities such as the
interfacial tension, free energy, and particle configurations computed with
our new theoretical framework by the two step HML method introduced above in
order to demonstrate its performance. We observe that the interfacial tension
$\overline{\gamma}^{ex}(\phi,c)$ decreases with increasing particle density
$c$, see \reff{Ge1}, and also a non-uniform interfacial energy $e(\phi)$ due
to the presence of particles on the interface and the dependence of phase
field parameter $\phi$ on $c$. Herewith, we have a powerful tool which
provides new research directions such as designing specific interfacial
geometries by adding new interfaces at locations where the energy density is
low.

In our second computation we consider $350$ particles which diffuse in one phase (blue part in Fig. 4) 
of the domain $[-1,1]^2$. We observe that our computational framework
captures the experimentally observed decrease of interfacial energy (Fig.
3(a) and Fig. 4(a)) by an increase of the interfacial particle density. This
demonstrates that we are able to compute the dynamic surface tension which
depends on the interfacial particle density and geometry. Up to now, there
was has not been a formalism which considers the crucial influence of the
interfacial geometry. We emphasize that a further novelty of our framework is
that for dynamic interfaces we account for the wetting characteristics of the
particles by the contact angle which determines the adsorption rate $\alpha$,
see \reff{fI} and \reff{padsorb}. As our methodology is based on a new free
energy that incorporates all fundamental ingredients of particles' adsorption
on dynamic interfaces, we can compute any physical quantity of experimental
interest, such as the influence of the interfacial geometry and the
particles' wetting properties on the adsorption rate, for instance. Another
possibility would be velocity profiles depending on particle density or the
critical particle density for drop-drop coalescence and drop breakup.

\begin{figure}[ht]
\includegraphics[width=0.2\textwidth]{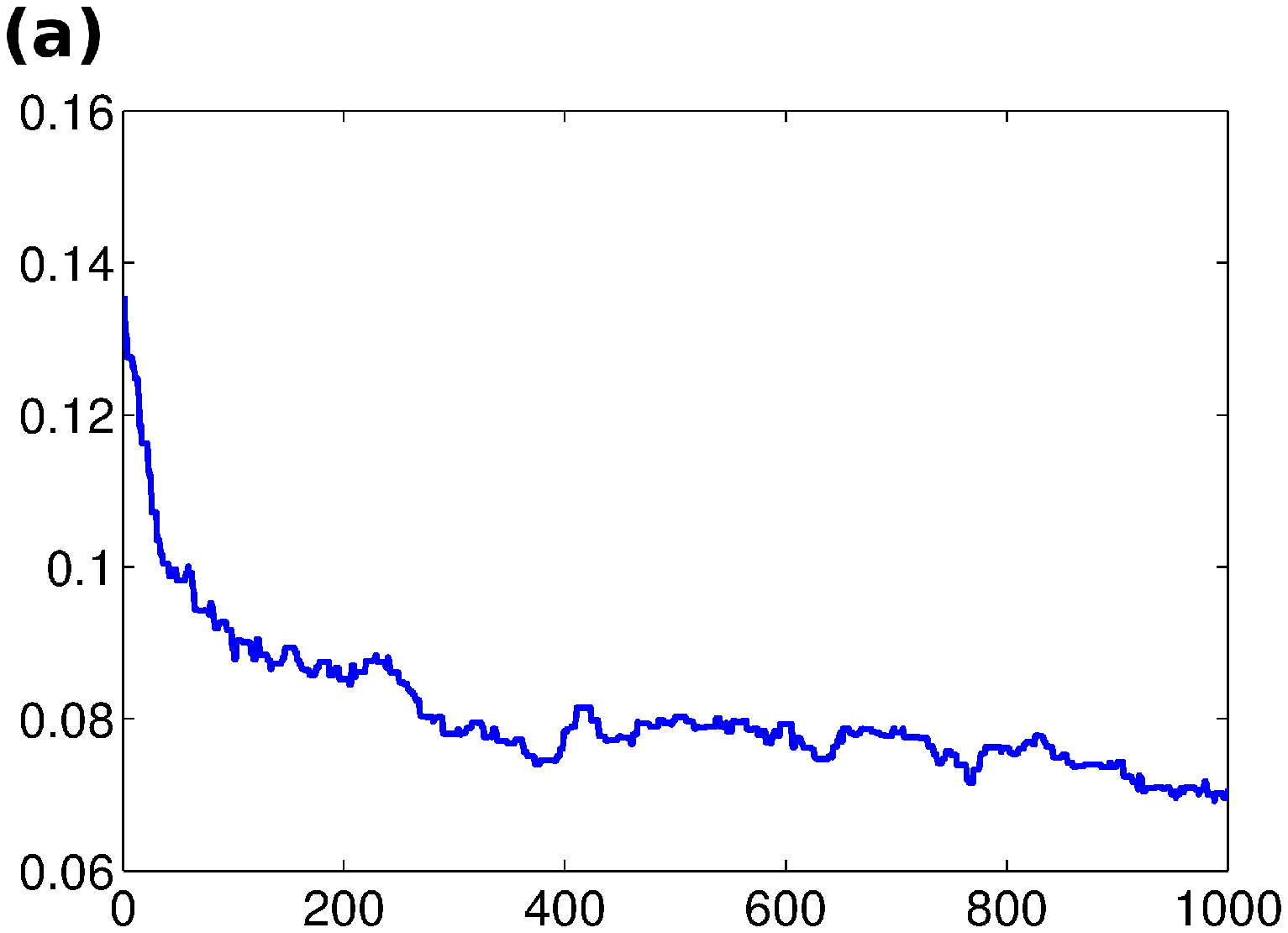} 
\hfill
\includegraphics[width=3.2cm]{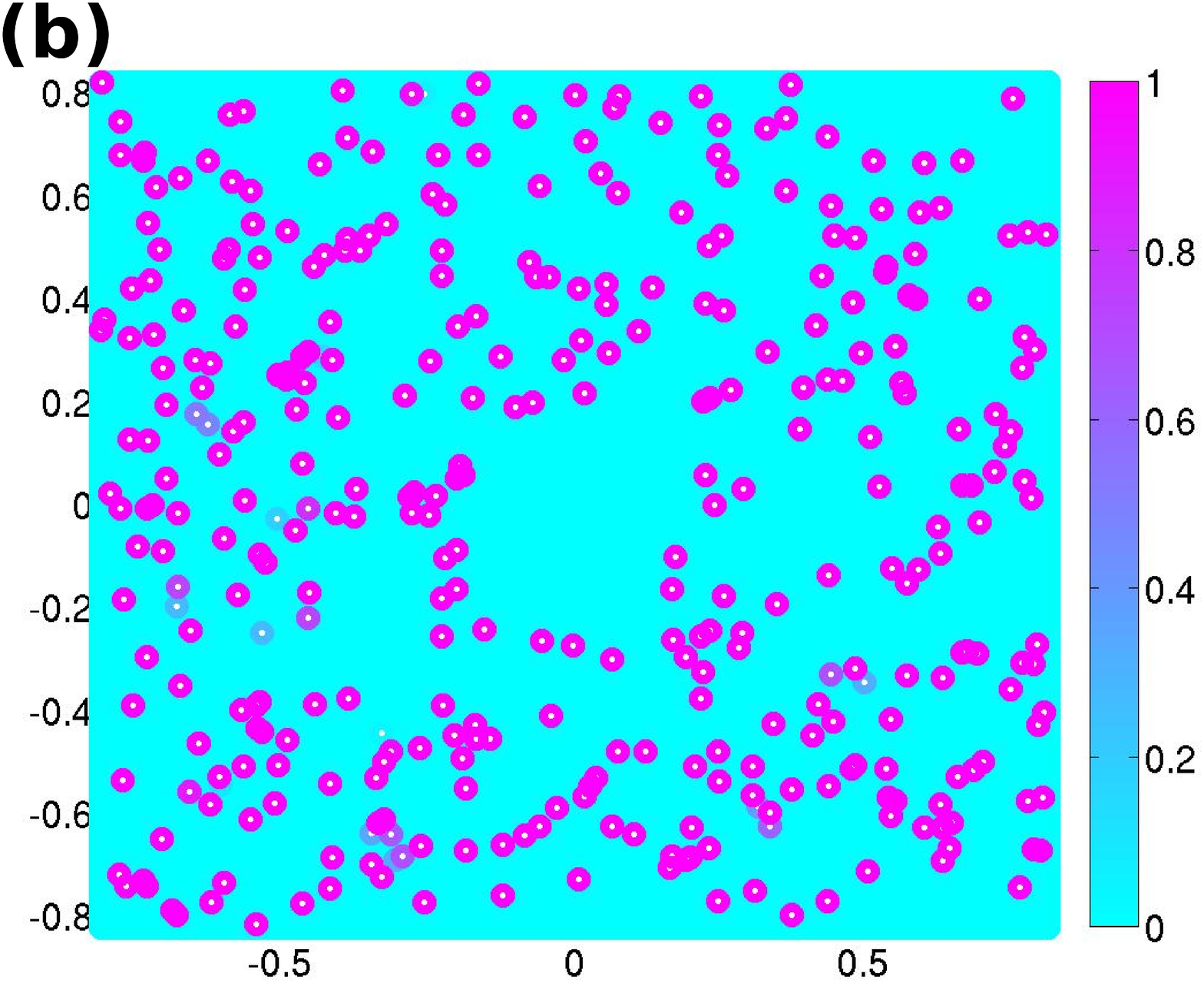}
\\
\includegraphics[width=3.5cm]{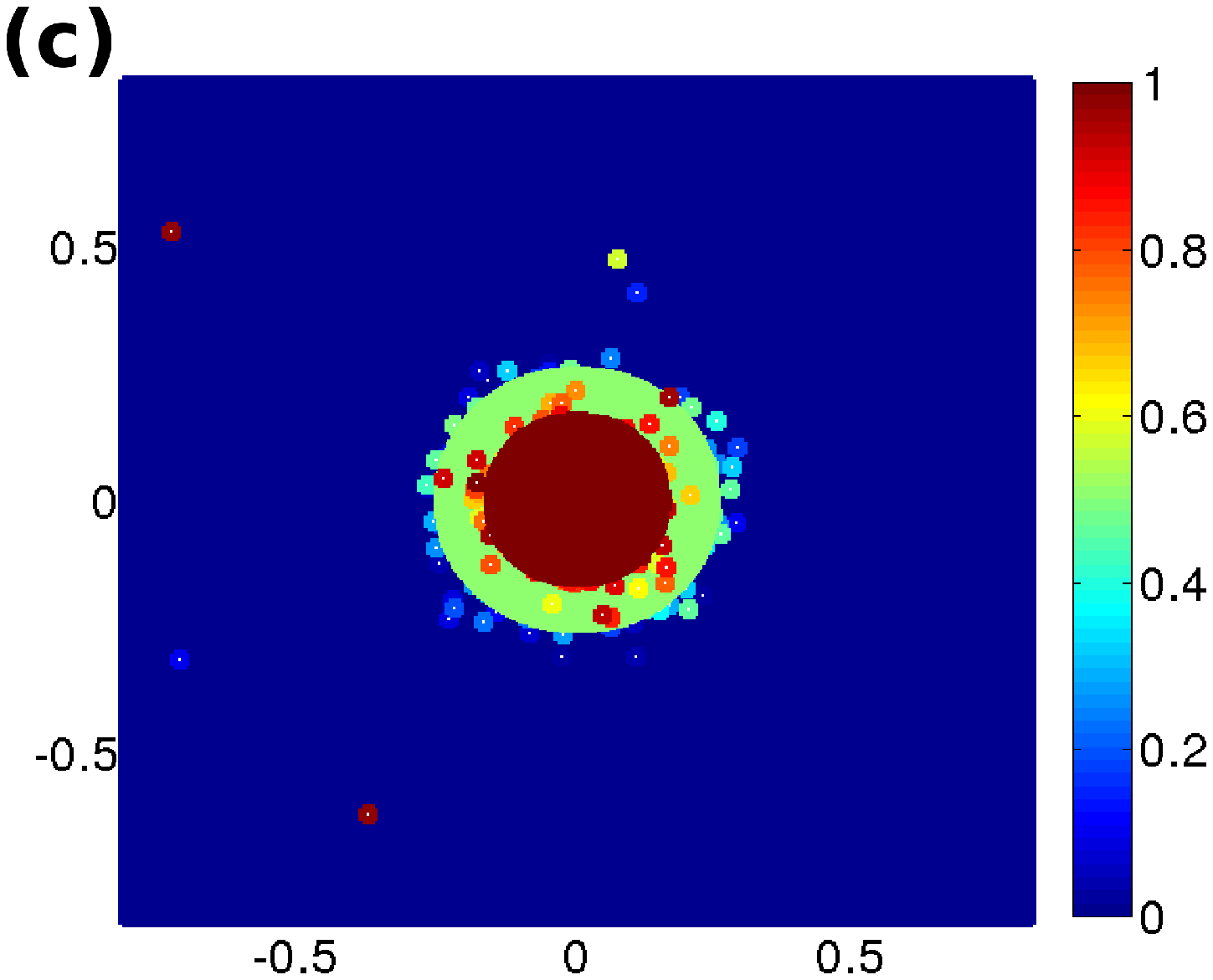}
\hfill
\includegraphics[width=3.5cm]{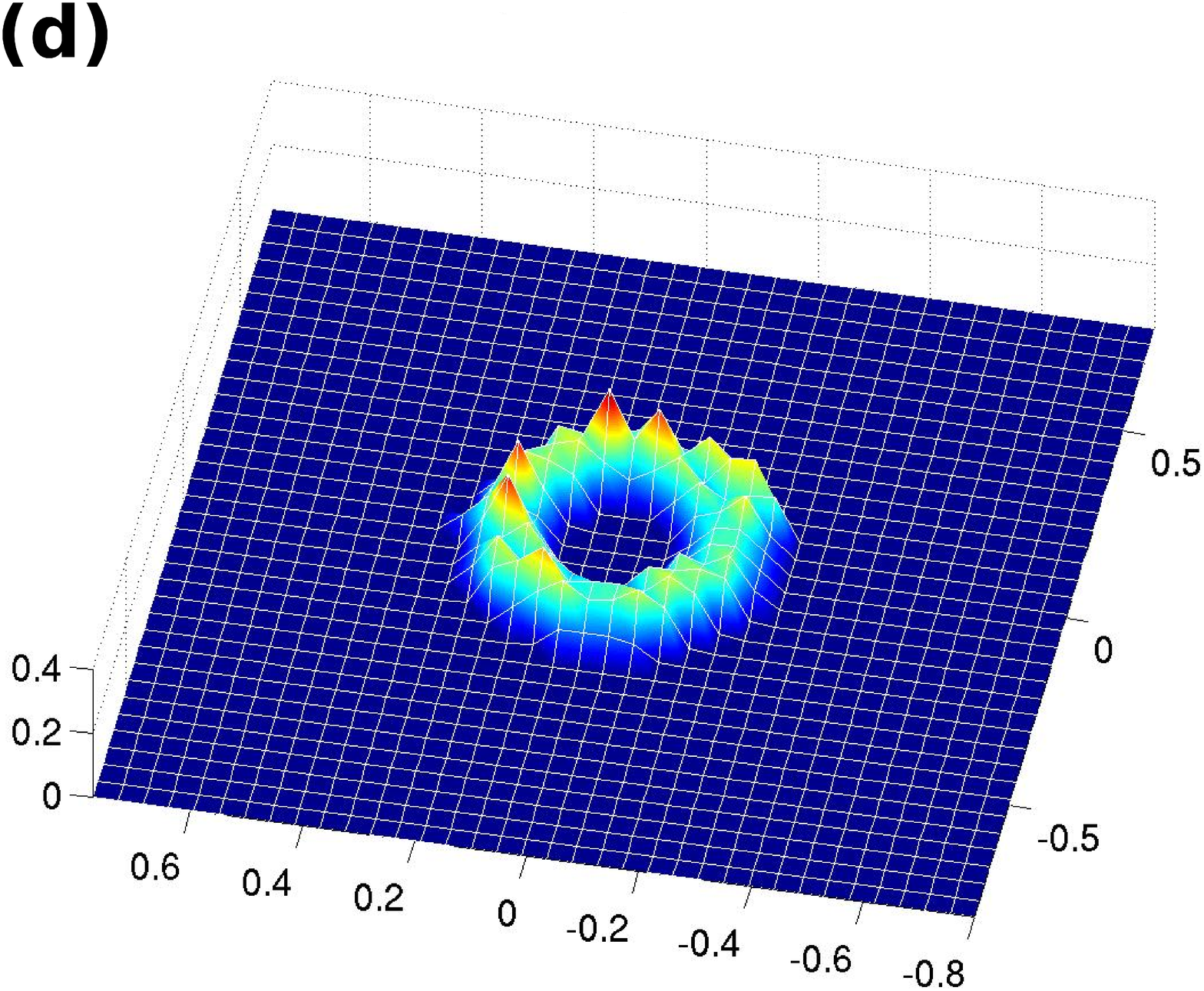}
\caption{{\bf (a)} Decay of the interfacial tension by minimizing the free energy \reff{Ge1} via a stochastic HML method.
{\bf (b)} Microscopic particle dynamics represented by $350$ particles.
{\bf (c)} Microscopic computation of the phase field parameter $\phi$. {\bf (d)} Interfacial energy.
}
\label{figPhQ3}
\end{figure}


\section{Conclusion}
\label{sec:4}
We have outlined a new and general framework for particle adsorption on
immiscible fluid interfaces by a thermodynamic free energy formulation
leading to a new coupled system of PDEs. We further proposed a new efficient
multi-level discretization strategy which accounts for the multi-scale
feature of this complex multi-phase problem characterized by the presence of
different length- and time-scales. Our formulation incorporates the
characterization of the particles by the contact angle and captures the
coupled-nonlinear dynamics between particle density and the interfacial
location-geometry. The generality of our framework also enables to model
surfactants, the incorporation of particles with non-local
interactions~\cite{Tcholakova2008,Zeng2012} or even of free energies obtained
from first principles, such as density functional theory for instance,
e.g.~\cite{Evans99,Yatsyshin2012,Ben2012,Yatsyshin2013}, while of particular
interest would be the extension of the
framework to wetting problems,
e.g.~\cite{Eggers2009,Savva2010,Savva2011}.

The strength of our methodology relies on its variational structure which
allows to derive analytical expressions such as adsorption isotherms at
thermodynamic equilibrium. These isotherms reduce in specific limits to the
classical Langmuir isotherm for instance and generalize existing ones by
taking the interfacial location and geometry into account. In addition to the
agreement of a decreasing interfacial tension by an increasing particle
density between experiments \cite{Garbin2012b} and our computations (Fig. 3
and Fig. 4), we demonstrated the generality and flexibility of our framework
by studying systems of mixed particles and even systems with fluid flow. This
is a promising direction towards designing fluids with highly tunable
transport properties. For instance, one could characterize conditions (e.g.
fluid velocity, particle density, pressure, temperature, etc.) under which
the particles are trapped on the interface and hence be transported between
two locations. This should be of relevance in applications such as
microfluidics and drug delivery. We also demonstrated the numerical
efficiency of our framework with computations based on the new HML method.

These results offer a solid basis for the simulation and control of particle
adsorption in a wide variety of physical phenomena and technological
problems. The compatibility of our methodology with first-principle modelling
strategies together with the scale resolving multi-level discretization
approach, provides a promising tool to computationally exploiting complex
particulate multiphase systems and their rich nature.





\bibliographystyle{elsarticle-num}
\bibliography{padsorpCH_main.bib}







\end{document}